\documentclass[twocolumn,aps,prb,showpacs,superscriptaddress,amsfonts,amssymb,amsmath]{revtex4}

\usepackage{graphicx}
\begin{document}

\title{Quantum transport theory for nanostructures with Rashba spin-orbital
interaction}

\author{Qing-feng Sun$^{\ast}$\\
 {\sl\small $^1$Department of Physics, The University of Hong
Kong, Pokfulam Road, Hong Kong, China;\\
$^2$Beijing National Lab for Condensed Matter Physics and
Institute of Physics, Chinese Academy of Sciences, Beijing 100080,
China \\ }
{ Jian Wang}\\
{\sl\small
 Department of Physics, The
University of Hong Kong, Pokfulam Road, Hong Kong, China} \\
{ Hong Guo}\\
{\sl\small Center for the Physics of Materials and Department of
Physics, McGill University, Montreal, PQ, Canada H3A 2T8}}

\date{\today}

\begin{abstract}
We report on a general theory for analyzing
quantum transport through devices in the Metal-QD-Metal
configuration where QD is a quantum dot or the device scattering
region which contains Rashba spin-orbital and electron-electron
interactions. The metal leads may or may not be ferromagnetic,
they are assumed to weakly couple to the QD region. Our theory is
formulated by second quantizing the Rashba spin-orbital
interaction in spectral space (instead of real space), and quantum
transport is then analyzed within the Keldysh nonequilibrium
Green's function formalism. The Rashba interaction causes two main
effects to the Hamiltonian: (i) it gives rise to an extra
spin-dependent phase factor in the coupling matrix elements
between the leads and the QD; (ii) it gives rise to an inter-level
spin-flip term but forbids any intra-level spin-flips. Our
formalism provides a starting point for analyzing many quantum
transport issues where spin-orbital effects are important. As an
example, we investigate transport properties of a Aharnov-Bohm
ring in which a QD having Rashba spin-orbital and e-e interactions
is located in one arm of the ring. A substantial spin-polarized
conductance or current emerges in this device due to a combined
effect of a magnetic flux and the Rashba interaction. The
direction and strength of the spin-polarization are shown to be
controllable by both the magnetic flux and a gate voltage.
\end{abstract}


\maketitle

\section{Introduction}
\label{sec1}

The spin-orbit (SO) interaction in semiconductors has attracted
great attention in recent
years\cite{ref1,ref2,ref3,ref4,addref1,ref5,ref6,ref7,ref8,ref9,addref2}
as it plays a very interesting role for the emerging field of
semiconductor spintronics. SO interaction can couple the spin
degree of freedom of an electron to its orbital motion and vice
versa thereby giving a useful handle for manipulating and
controlling the electron spin by external electric fields or gate
voltages. SO is an intrinsic interaction having its origin from
relativistic effects, but it is believed to be substantial in some
semiconductors. More than ten years ago, Datta and Das
theoretically analyzed the possibility of a spin-transistor which
works due to Rashba SO interaction that induces spin precessions
in a semiconductor\cite{ref6} with ferromagnetic leads. More
recently, Murakami {\it et.al.} theoretically predicted that a
substantial amount of dissipationless quantum spin current could
be generated by a co-action of electric field and SO
interaction\cite{ref4,addref1}. Shen {\it et.al.} found a resonant
spin Hall conductance in a two-dimensional (2D) system with Rashba
SO interaction under a perpendicular magnetic field\cite{ref5}.
There are also many other works on related issues where SO
interaction plays a central
role\cite{ref10,ref11,ref12,ref13,ref14,addref3}, and this
research direction is expanding in a very rapid pace due to its
possible application to spintronics.

A semiconductor spintronic device is likely to be based on
mesoscopic and nanostructures where electron-electron (e-e)
interactions may be strong. Hence it is desirable to formulate a
general quantum transport theory which can handle SO, e-e, and
other interactions for systems in the Metal-QD-Metal
configuration. Here the ``QD'' indicates a quantum dot or the
device scattering region where the various interactions exist,
while the ``Metals'' are device leads which extend to electron
reservoirs far away. The metal leads may or may not be
ferromagnetic, but are weakly coupled to the QD region. In almost
all previous theoretical work, the SO interactions are represented
by a real space Hamiltonian where e-e interactions and strong
correlations are usually neglected. Indeed, it is rather
difficult, if not impossible, to handle SO together with e-e
correlations and other interactions in real space for transport
problems. In contrast, the most powerful and general theoretical
technique for quantum transport in meso- and nano-scopic systems
is the Keldysh nonequilibrium Green's function (NEGF)
formalism\cite{rammer-book}. NEGF can handle many-body
correlations and interactions in an unified fashion and is a well
established formalism\cite{rammer-book}. NEGF is typically
formulated in momentum space or other spectral space for
theoretical and numerical analysis. This means that all
interactions need to be formulated in the spectral space. In other
words, in order to conveniently apply NEGF theory, one needs to
write the SO interactions in a spectral space with second
quantization. To the best of our knowledge, we are not aware of a
derivation of such a second quantized form for SO interaction.

It is the purpose of this paper to report a general quantum
transport theory for Metal-QD-Metal devices with SO and e-e
interactions, based on the NEGF framework. We begin by presenting
a pedagogical discussion of the general physics of SO interaction
by quantizing the corresponding {\it classical} action, which
gives a vivid physical picture of SO interaction. We then second
quantize the real space SO interaction Hamiltonian in a proper
spectral space so that it can be analyzed by NEGF for quantum
transport properties of Metal-QD-Metal devices. Although the
derivations are general, we specialize on a Metal-QD-Metal device
where the QD is described by the Anderson model plus the Rashba SO
interaction, and the leads are ferromagnetic metal. The second
quantized Hamiltonian can then be analyzed within NEGF and well
established many-body theoretical
methods\cite{ref15,ref16,ref17,ref18}. To illustrate our
formalism, we investigate quantum transport properties of a
Aharnov-Bohm ring where a QD having Rashba SO and e-e interactions
sits in one arm of the ring. We found that a substantial
spin-polarized conductance or current emerges in this device when
a magnetic flux passes through the ring. In particular, its
spin-polarized direction and strength are controllable by both the
magnetic flux and a gate voltage, hence the predictions are
testable experimentally.

The paper is organized as follows. In the next section we discuss
the many-body Hamiltonian of a general Metal-QD-Metal device
structure, and present a pedagogical discussion of the SO
interactions in real space. We then proceed, in Section III, to
second quantize the Rashba SO interaction in spectral space so
that the entire device Hamiltonian can be second quantized. This
process is divided into several subsections and careful
derivations and discussions will be presented. A brief summary for
the second quantized Hamiltonian in spectral space is given in
Section II-F. In section IV, we analyze quantum transport
properties of a modified AB ring which contains a QD in one arm of
the ring, and both Rashba SO and e-e interactions exist in the QD.
Finally, Section V summarizes the results of our work.

\section{Hamiltonian of the Metal-QD-Metal device}
\label{sec2}

In this section we discuss the Hamiltonian of a general device
structure in the Metal-QD-Metal device configuration. By
presenting a very useful pedagogical discussion on the classical
forces acting on moving charges and spins inside electrical and
magnetic fields, we realize that the SO interaction originates
from the force (torque) of electrical field on the moving spin.
This allows us to write down the Hamiltonian for the SO
interaction in real space, $H_{so}({\bf r})$, and in particular we
derive the Rashba SO interaction. Of course, the derived
$H_{so}({\bf r})$ is the same as that from the Dirac
equation\cite{bjorken}, but the pedagogical discussion gives a
vivid physical picture about SO interactions for quantum transport
in solid state devices. In fact, in the literature of relativistic
quantum mechanics such as the book of Bjorken and
Drell\cite{bjorken}, SO interaction has been discussed with the
point of view of quantizing the classical force acting on the
moving spin by the external electric field. We found such a
pedagogical discussion in the context of solid state electronics
to be very useful.

The general device structure we consider is schematically shown in
Fig.1a, where the scattering region (QD) is connected to the
outside world by coupling to two ferromagnetic metal leads (FM).
The permanent magnetic moments of the leads are denoted by vectors
${\bf M}_\beta$ where $\beta=L,R$ indicate the left and right
leads. The QD is assumed to be weakly coupled to the leads due to
the potential barriers at the two Metal-QD junctions (Fig.1b).
Inside the QD there are SO and e-e interactions, while these
interactions are neglected in the leads. There may also be an
external magnetic field ${\bf B}({\bf r})$. For this device, the
total many-body Hamiltonian can be written as
\begin{equation}
 H({\bf r}_1, {\bf r}_2, ..., {\bf r}_N) =
 \sum\limits_i H_s({\bf r_i})
  + \sum\limits_{i,j (i\not=j)} H_I({\bf r}_i,{\bf r}_j)
\label{Hdevice}
\end{equation}
where the second term is the e-e interaction $H_I({\bf r}_1,{\bf
r}_2)$ and the first term is from the single particle Hamiltonian
$H_s({\bf r})$:
\begin{equation}
 H_I({\bf r}_1,{\bf r}_2) = \frac{e^2}{2|{\bf r}_1-{\bf r}_2|^2}
\end{equation}
\begin{eqnarray}
 H_s({\bf r}) & = & \frac{{\bf p}^2}{2m^*} + V({\bf r})
 + \hat{\bf
\sigma} \bullet {\bf M}({\bf r})
+  \hat{\bf \sigma} \bullet {\bf B}({\bf r})  \nonumber \\
& & + H_{so}({\bf r})\ \ . \label{Hs1}
\end{eqnarray}
$H_s$ contains the usual single particle terms: the kinetic and
potential energies, the interaction energy with the magnetic
moment ${\bf M}$ in the ferromagnetic leads, and the Zeeman
energy. The last term in Eq(\ref{Hs1}) is the SO interaction
$H_{so}$. Although the real space form of $H_{so}$ is
known\cite{ref19,ref20} from Dirac equation, in the following we
present a pedagogical discussion about it.

Transport in our device is about the motion of two
entities---charge and spin, in two fields---electric and magnetic
fields. Therefore there are a total of four actions due to the
fields on the charge/spin: (i) the electric or Coulomb force on
charge; (ii) the Lorentz force on moving charge; (iii) the
magnetic force on spin (Zeeman); and (iv) the electric force on
moving spin. Of these four actions, (i)-(iii) are well known and
familiar, but (iv) is much less so. Where does (iv) comes from? It
comes due a purely relativistic effect\cite{ref21}. Consider a
spin which produces a magnetic field in the space surrounding it:
if this spin is moving, by relativistic transform we obtain an
electric field (in addition to the magnetic field). In other
words, a moving spin produces an electric field. Conversely, if a
spin is moving inside an {\it external} electric field, it will be
subjected to an action (torque). In this sense, (iv) is the
counter part of the Lorentz force. It has been shown that a moving
spin with velocity {\bf v} inside an electric field {\bf E} is
subjected to a torque action with an interactive potential
energy:\cite{ref21,addbook}
\begin{equation}
\frac{ e\hbar}{4m_e c^2} {\bf \sigma}\bullet({\bf v}\times{\bf E})
\label{torque1}
\end{equation}
where $m_e$ is the electron mass, $c$ the speed of light, and the
electron spin is ${\bf s}=\frac{\hbar}{2}{\bf \sigma}$. Of course,
this is a classical expression.

To quantize the classical torque energy (\ref{torque1}), we make
the following replacements: electric field ${\bf E} \rightarrow
\nabla V({\bf r})/e$ where $V({\bf r})$ is the potential energy of
the system; the speed ${\bf v} \rightarrow {\bf p}/m_e$ where
${\bf p}$ is the momentum operator. The quantum mechanical
correspondence of expression (\ref{torque1}) becomes:
\begin{eqnarray}
& & \frac{\hbar}{8m_e^2 c^2 }
   \left[ \hat{\bf \sigma} \bullet ({\bf p} \times \nabla V({\bf r}))
       -  \hat{\bf \sigma} \bullet (\nabla V({\bf r})\times {\bf p} )
   \right] \nonumber\\
 &= &
 \frac{\hbar}{8m_e^2 c^2 }
   \left[ \hat{\bf \sigma} \bullet ({\bf p} \times \nabla V({\bf r}))
       +  \nabla V({\bf r}) \bullet (\hat{\bf \sigma}\times {\bf p} )
   \right]
\label{Hso1}
\end{eqnarray}
where $\hat{\bf \sigma} =(\hat{\sigma}_x, \hat{\sigma}_y, \hat{\sigma}_z)$
is the vector of Pauli matrix. Expression (\ref{Hso1}) is exactly the general
form of the SO interaction Hamiltonian, usually derived from the Dirac
equation
in the low speed limits\cite{bjorken}. Therefore, the essence of the SO
interaction is simply the action of an external electric field on a moving
spin.

If the potential $V({\bf r})$ has spherical symmetric, {\it i.e.}
$V({\bf r}) = V(r)$, we have $\nabla V({\bf r}) = \frac{\bf
r}{r}\frac{d}{dr}V(r)$. Then the general spin-orbit interaction of
Eq.(\ref{Hso1}) reduces to the following familiar form:
\begin{equation}
 H_{so} = -\frac{1}{2 m_e^2 c^2} \frac{1}{r} \frac{d}{dr}V(r)
   \hat{\bf s} \bullet \hat{\bf l}
   \label{Hso2}
\end{equation}
where the orbital angular momentum operator is $\hat{\bf l} = {\bf
r}\times {\bf p}$.  In fact, Eq.(\ref{Hso2}) is the well-known
Thomas SO coupling.

When our device is made of a two-dimensional electron gas (2DEG)
in which electrons are strongly confined in the $y$ direction by a
confining potential $V(y)$, such that $\frac{dV}{dy} \gg
\frac{dV}{dx}, \frac{dV}{dz}$, then $\nabla V({\bf r}) \approx
\hat{y} \frac{dV}{dy} $ and the electric field is almost along the
$y$ direction. Furthermore, if $V(y)$ is asymmetric with respect
to the reflection point $y=0$, then the matrix element $\langle
\Psi(y)|\frac{d}{dy} V(y)|\Psi(y)\rangle \neq 0$ where $\Psi(y)$
is the basic bound states in the $y$ direction. Under these
conditions, the general SO interaction Eq.(\ref{Hso1}) reduces to
the Rashba SO interaction form:\cite{ref19,ref20}
\begin{equation}
H_{so}
 = \frac{\hat{y}}{2\hbar} \bullet \left[
        \alpha(\hat{\bf \sigma} \times {\bf p}) +
       (\hat{\bf \sigma} \times {\bf p}) \alpha \right]
\label{Rashba1}
\end{equation}
where $\alpha \sim \langle \Psi(y)|\frac{d}{dy} V(y)|\Psi(y)
\rangle$ is the interaction coefficient. Note that an asymmetrical
confining potential in the direction perpendicular to the 2DEG
(the $\hat{{\bf y}}$ direction) is necessary, otherwise $\alpha
=0$ and there will be no Rashba SO interaction. It is worth to
mention that the Rashba SO interaction strength $\alpha$ can be
tuned in an experiment by an external electric field or gate
voltage, which have already been done in some recent
experiments\cite{ref22,ref23,ref24,ref25}. Finally, if we consider
other forms of the potential energy $V({\bf r})$, we obtain other
kinds of SO interactions, but the essence of SO coupling is the
interaction of the external electric field on the moving spins.

\section{Second quantization of the device Hamiltonian}
\label{sec3}

In this section we second quantize the device Hamiltonian (\ref{Hdevice}). The
focus is to derive the second quantization of the Rashba SO interaction
in a spectral form.

\subsection{Without SO interaction}
\label{sec3.1}

The second quantized form for the Hamiltonian (\ref{Hdevice}) of
the Metal-QD-Metal device with non-magnetic leads (${\bf M}=0$),
in zero magnetic field (${\bf B}=0$), and without SO interaction
($\alpha=0$) can be approximately written in the standard Anderson
model:
\begin{equation}
 H\ =\ H_{QD} + \sum_{\beta=L,R} H_\beta + H_T
\end{equation}
where $H_{QD}$ is the Hamiltonian for the QD region; $H_\beta$ is
for the leads and $H_T$ the coupling between the leads and the QD:
\begin{eqnarray}
& & \hspace{-0.8cm}H_{QD}  =
      \sum\limits_{n,s}\epsilon_{n} d^{\dagger}_{ns} d_{ns}
     +\sum\limits_{ns,ms'(ns\not= ms')} U_{ns,ns'} \hat{n}_{ns}\hat{n}_{ms'}
\label{HQD1} \\
& & \hspace{-0.8cm}H_{\beta} = \sum\limits_{k,s} \epsilon_{k\beta}
     a^{\dagger}_{k\beta s} a_{k\beta s}
\label{Hlead1} \\
& &\hspace{-0.8cm} H_T  =  \sum\limits_{k \beta,n,s}
      \left[t_{k\beta n} a^{\dagger}_{k\beta s} d_{ns}
           +H.c \right]
\label{HT1}
\end{eqnarray}
The quantity $\hat{n}_{ns}\equiv d^{\dagger}_{ns} d_{ns}$;
$s=\uparrow, \downarrow$ (or + and -) is the spin index which also
describes the spin states, with $|s \rangle =(1,0)^T$ and
$(0,1)^T$ for the spin-up and spin-down states, respectively. $n$
is quantum number for eigenstates of the single-particle
Hamiltonian $H_s$ (Eq.(\ref{Hs1})) in the isolated QD region with
eigen-energy $\epsilon_n=\langle n|H_s|n\rangle$; $k\beta$ is the
quantum index for lead $\beta$ with eigen-energy
$\epsilon_{k\beta}=\langle k\beta|H_s|k\beta \rangle$
($\beta=L,R$). $t_{k\beta,n} =\langle k\beta|H_s|n \rangle$
describes coupling strength between the leads and the QD region.
Quantity $U_{ns,ns'} = \langle ns,ms'|\frac{e^2}{2|{\bf r}_1-{\bf
r}_2|^2} |ms',ns \rangle$ is the matrix element for the two-body
e-e interaction. Here the e-e interaction in the leads region has
been neglected. Note that when ${\bf M}={\bf B}=\alpha =0$, $H_s$
does not depend on the Pauli matrix $\hat{\bf \sigma}$, therefore
$[\hat{\bf \sigma}, H_s] =0$ and $\epsilon_n$,
$\epsilon_{k\beta}$, and $t_{k\beta,n}$ are all independent of the
spin-index $s$.

For ferromagnetic leads ${\bf M}_{\beta}\neq 0$ and $H_s$ of
Eq.(\ref{Hs1}) contains a term $\hat{\bf \sigma}\bullet{\bf
M}_{\beta}$. Let's assume ${\bf M}_{\beta}$ has a constant value
in each lead $\beta$ although ${\bf M}_L \neq {\bf M}_R$ in
general. By calculating matrix elements $\langle k\beta s|
\hat{\bf \sigma} \bullet {\bf M}_{\beta} |k'\beta s' \rangle =
\delta_{kk'}(\hat{\bf \sigma}\bullet {\bf M}_{\beta})_{ss'}$, the
second quantization for this term can be easily obtained.
$H_{\beta}$ becomes:
\begin{eqnarray}
  H_{\beta}  =  \sum\limits_{k,s} \epsilon_{k\beta}
     a^{\dagger}_{k\beta s} a_{k\beta s}
     + \sum\limits_k
     (a^{\dagger}_{k\beta\uparrow},a^{\dagger}_{k\beta\downarrow})
       \hat{\bf \sigma}\bullet{\bf M}_{\beta}
              \left( \begin{array}{l}
         a_{k\beta\uparrow}  \\
         a_{k\beta\downarrow}
         \end{array}\right)
\nonumber
\end{eqnarray}
Due to the existence of $\hat{\bf \sigma}\bullet {\bf M}_{\beta}$,
the state $|k\beta s \rangle$ is usually not an eigenstate of
isolated lead $\beta$, and $\langle a^{\dagger}_{k\beta s}
a_{k\beta s} \rangle $ is not equal to the Fermi distribution
function $f_{\beta}(\epsilon_{k\beta})$ even in equilibrium. In
order to conveniently solve the transport problem, we diagonalize
$H_{\beta}$ above by a rotational unitary transformation. This is
accomplished by setting:
\begin{equation}
 \left( \begin{array}{l}
    a_{k\beta \uparrow'} \\ a_{k\beta \downarrow'}
 \end{array} \right)
   =
 \left( \begin{array}{ll}
    \cos \frac{\theta_{\beta} }{2} e^{i\phi_{\beta}/2} &
    \sin \frac{\theta_{\beta}}{2} e^{-i\phi_{\beta}/2} \\
    -\sin \frac{\theta_{\beta}}{2} e^{i\phi_{\beta}/2} &
    \cos \frac{\theta_{\beta}}{2} e^{-i\phi_{\beta}/2}
 \end{array} \right)
 \left( \begin{array}{l}
    a_{k\beta\uparrow} \\ a_{k\beta\downarrow}
 \end{array} \right)
\nonumber
\end{equation}
where $\theta_{\beta}$ and $\phi_{\beta}$ are the directional angles
of the FM moment ${\bf M}_{\beta}$. Under this rotational transformation,
the total second quantized Hamiltonian of the Metal-QD-Metal device
becomes:
\begin{eqnarray}
& &  H_{QD}  = \sum\limits_{n,s}\epsilon_{n}
             d^{\dagger}_{ns} d_{ns}
     +\sum\limits_{ns,ms'(ns\not= ms')}
         U_{ns,ms'}  \hat{n}_{ns}\hat{n}_{ms'}
        \nonumber \\
&& H_{\beta} = \sum\limits_{ k,s} (\epsilon_{k\beta }
  +s  M_{\beta}) a^{\dagger}_{k\beta s} a_{k\beta s}
\label{H2} \\
&&  H_T = \sum\limits_{k \beta,n,s}
    \left[t_{k\beta n} \left( \cos \frac{\theta_{\beta}}{2}
            a^{\dagger}_{k\beta s} -s\sin \frac{\theta_{\beta}}{2}
            a^{\dagger}_{k\beta \bar{s}}\right) \times \right. \nonumber \\
&& \left.\hspace{2.0cm} e^{is\phi_{\beta}/2} d_{ns} +H.c \right]
\nonumber
\end{eqnarray}
where $M_{\beta}=|{\bf M}_{\beta}|$. This form of the Hamiltonian
has been used before\cite{ref26}, but two important comments need
to be made: (i) In Hamiltonian (\ref{H2}), the states $|k\beta s
\rangle $ are eigenstates of $H_{\beta}$ for isolated leads, hence
in equilibrium $\langle a^{\dagger}_{k\beta s} a_{k\beta s}
\rangle = f_{\beta}(\epsilon_{k\beta s})$ with $\epsilon_{k\beta
s} \equiv\epsilon_{k\beta}+sM_{\beta}$. (ii) After the rotational
transformation, the spin-up direction in the left FM lead, the QD
and the right FM lead, are all different although they are all
aligned in their local $\hat{\bf z}$-direction. These local
coordinate systems are shown in Fig.2. In the QD, the spin-up
direction is still in the original $\hat{\bf z}$ axis; but in the
left/right FM leads, the spin-up direction ({\it i.e.} local
$\hat{\bf z}$-direction) is aligned with the FM moment ${\bf
M}_{L/R}$ (see Fig.2). Although this difference of spin-up
alignment is not important when the QD bridging the leads has only
a single connection (such as Fig.1), it is important if the QD
region has double or more connections (such as Fig.5).

\subsection{Rashba SO interaction (I)}

In this and the next subsections, we second quantize the Rashba SO
interaction which is a major content of this paper. The Rashba SO
interaction (\ref{Rashba1}) can be splited into two terms:
\begin{eqnarray}
 \frac{\hat{y}}{2\hbar} & \bullet & \left[
        \alpha(x)(\hat{\sigma} \times {\bf p})  +
       (\hat{\sigma} \times {\bf p}) \alpha(x) \right] \nonumber \\
 &=&
 \frac{1}{2\hbar}  \left[
        \alpha(x) \hat{\sigma}_z  p_x  +
      \hat{\sigma}_z  p_x \alpha(x) \right]
    -\frac{\alpha(x) \hat{\sigma}_x p_z}{\hbar} \nonumber \\
 &\equiv &H_{R1} +H_{R2}\ \ .
\label{Rashba2}
\end{eqnarray}
For transport direction along the $\hat{\bf x}$ axis as shown in Fig.1,
these two terms have some essential differences. The first term $H_{R1}$
gives rise to a spin precession\cite{ref6} while the second term $H_{R2}$
does not. In particular, $H_{R1}$ includes a $\delta$-function factor
at the Metal-QD contacts ($x=x_{L/R}$, see Fig.1) with $\alpha=0$ and
the QD region with $\alpha \neq 0$\cite{ref27}. For this reason
it cannot be second quantized by simply calculating the matrix element
$\langle ns|H_{R1}|ms' \rangle $. To overcome this difficulty
one has to choose a new basis set in the QD. This will be accomplished
in this subsection and the $H_{R2}$ term will be studied in the next
subsection.

For clarity, the real space single particle Hamiltonian considered
in this subsection is:
\begin{eqnarray}
H_s^1({\bf r}) \equiv \frac{p^2_x +p_z^2}{2m^*} +V({\bf r})
                 +\hat{\bf \sigma}\bullet {\bf M}({\bf r}) + H_{R1}\ .
\label{Hs1R1}
\end{eqnarray}
This is just Eq.(\ref{Hs1}) but with only the $H_{R1}$ part of the
SO interactions. We make an unitary transformation with the
following unitary matrix:
\begin{eqnarray}
  u(x) = \left\{
  \begin{array}{ll}
    1 &  x<x_L \\
    exp\{-i\hat{\sigma}_z\int^x_{x_L} k_R(x) dx \}& x_L <x<x_R \\
    exp\{-i\hat{\sigma}_z\int^{x_R}_{x_L} k_R(x) dx\} & x_R <x
  \end{array}\right.
\label{U1}
\end{eqnarray}
where $k_R(x) \equiv {\alpha(x)m^*/\hbar^2}$. Here $\alpha(x)$ is
permitted to have a dependence on spatial coordinate $x$ inside
the QD, and it is zero outside ($x_R < x$ or $x < x_L$, see
Fig.1). Under this unitary transformation, the original basis
functions in the QD region, $|n\uparrow \rangle =\varphi_n({\bf
r})(1,0)^T$ and $|n\downarrow \rangle =\varphi^*_n({\bf
r})(0,1)^T$, are transformed to
\begin{eqnarray}
|n\uparrow \rangle '& =& u(x)|n\uparrow \rangle  =
  e^{-i\int^x_{x_L} k_R(x) dx} \varphi_n({\bf r})
   \left( \begin{array}{l} 1\\0 \end{array}\right)
\label{bas1} \\
|n\downarrow \rangle ' &=& u(x)|n\downarrow \rangle  =
  e^{+i\int^x_{x_L} k_R(x) dx} \varphi^*_n({\bf r})
   \left( \begin{array}{l} 0\\1 \end{array}\right)
\label{bas2}
\end{eqnarray}
These new basis functions are used to second quantize (\ref{Hs1R1}).
After the unitary transformation, $H_s^1$ of Eq.(\ref{Hs1R1}) becomes:
\begin{eqnarray}
H_s^{1'} &= & u(x)^{\dagger}H_s^1 u(x)  \nonumber \\
  & = &
  \frac{p^2_x +p_z^2}{2m^*} +V({\bf r})
  -\frac{\hbar^2 k^2_R(x)}{2m^*}
                 +{\bf \sigma}\bullet {\bf M}'({\bf r})
\label{Hs1p}
\end{eqnarray}
where ${\bf M}'_L = {\bf M}_L$ and $|{\bf M}'_R| = |{\bf M}_R|$,
but the directional angles of ${\bf M}'_R$ are changed to ($\theta_R$,
$\phi_R-2\phi_{so}$) with $\phi_{so}\equiv \int^{x_R}_{x_L} k_R(x)
dx$.

The essence of the above unitary transformation are the following.
(i) It is equivalent to choosing a space-dependent spin coordinate
as shown in Fig.3a, in which the spin $\hat{\bf z}$ direction is
fixed everywhere but the spin $\hat{\bf x}$ and $\hat{\bf y}$
directions are dependent on the space position ${\bf r}$. In
different positions along the ${\bf x}$ axis, the directions of
the spin $\hat{\bf x}, \hat{\bf y}$ axis are rotated. In other
words, the unitary transform changes us to a rotating frame. It is
well-known that for an electron moving along the ${\bf x}$
direction, the Rashba term $H_{R1}$ gives rise to a spin
precession\cite{ref6,ref13}. That is, the spin component in the
x-y plane will rotate as the electron moves along the $\hat{\bf
x}$ direction, therefore the electron spin is usually not
invariant. However, in the rotating frame which follows the spin
precession, the spin is invariant hence
$[H_s^{1'},\hat{\sigma}_{x/y/z}]=0$ is satisfied in the QD region.
(ii) The Rashba interaction $H_{R1}$ can cause an energy split
between spin-up and spin-down states for non-zero $k_x$, as shown
by the energy dispersion in the left panel of
Fig.3b\cite{ref16,ref28}. The above unitary transformation
recovers the alignment of the two dispersion curves so that the
right panel of Fig.3b is obtained. Therefore, after the unitary
transformation, the new Hamilton $H^{1'}_s$ appears to be
completely the same as the Hamiltonian without the Rashba
interaction $H_{R1}$, except a rotation of the magnetic moment
${\bf M}_R$ and a potential energy difference $-\frac{\hbar^2
k^2_R(x)}{2m^*}=-m^*\alpha^2(x)/(2\hbar^2)$ which is a simple
constant if $\alpha(x)$ is independent of $x$. Using the same
method as that of the last subsection, the second quantization of
Eq.(\ref{Hs1p}) is easily obtained:
\begin{eqnarray}
&& H  = H_{QD} + \sum_{\beta=L,R}H_{\beta} + H_T \nonumber \\
&& H_{QD}  = \sum\limits_{n,s}\epsilon_{n} d^{\dagger}_{ns} d_{ns}
        \nonumber \\
&&H_{\beta}  =  \sum\limits_{ k,s} (\epsilon_{k\beta} +sM_{\beta})
a^{\dagger}_{k\beta s} a_{k\beta s}  \\
&& H_T  =  \sum\limits_{k,n,s}
\left[t_{kL n} (\cos \frac{\theta_{L}}{2} a^{\dagger}_{kL s}
-s\sin \frac{\theta_{L}}{2} a^{\dagger}_{kL
\bar{s}})e^{is\phi_{L}/2} d_{ns}
\right.  \nonumber \\
&& + \left. t_{kR n} (\cos \frac{\theta_{R}}{2} a^{\dagger}_{kR s}
    -s\sin \frac{\theta_{R}}{2} a^{\dagger}_{kR\bar{s}})e^{is\phi_{R}/2}
        e^{-is\phi_{so}}d_{ns} \right. \nonumber \\
&& \left. \ \ \ \ \ \ \ \ \  +H.c \right]
\label{Hresult1}
\end{eqnarray}
This is one of the main results of this paper. Here the Rashba
interaction $H_{R1}$ gives rise to an extra spin dependent phase
factor $-s\phi_{so}$ in (\ref{Hresult1}): it is $-\phi_{so}$ for
$s=\uparrow$, and is $+\phi_{so}$ for $s=\downarrow$. Note that
the term with this phase factor satisfies the time-reversal
invariance while ${\bf M}_{\beta}=0$, {\it i.e.} $[T, H_{T}]=0$
where $T$ is the time-reversal operator.\cite{addnote} This is an
expected property because the Rashba SO interaction in real space
(Eq.(\ref{Rashba1})) does satisfy time-reversal invariance (see
the Appendix). We emphasize that the phase factor $-s\phi_{so}$ in
Eq.(\ref{Hresult1}) is fundamentally different from the phase
factor caused by magnetic flux in systems such as the AB-ring: the
latter is independent of spin $s$ and it destroys time-reversal
symmetry.

For the special case where $k_R(x)$=$k_R$=constant, {\it i.e.}
independent of coordinate $x$ of the scattering region, we have
$\phi_{so}= k_R\times (x_R-x_L)$. Then, re-defining $e^{-isk_R
x_L}d_{ns}\rightarrow d_{ns}$, the real space Hamiltonian
(\ref{Hresult1}) can be rewritten in a symmetric manner:
\begin{eqnarray}
  H_{QD} & = &
      \sum\limits_{n,s}\epsilon_{n}
             d^{\dagger}_{ns} d_{ns}
        \nonumber \\
  H_{\beta} & = & \sum\limits_{ k,s} (\epsilon_{k\beta} +sM_{\beta})
     a^{\dagger}_{k\beta s} a_{k\beta s} \label{Hresult2} \\
  H_T & =&  \sum\limits_{k,n,s,\beta}
      \left[t_{k\beta n}
   (\cos \frac{\theta_{\beta}}{2} a^{\dagger}_{k\beta s}
    -s\sin \frac{\theta_{\beta}}{2} a^{\dagger}_{k\beta\bar{s}})
    \right.
    \nonumber\\
   && \left. \times
    e^{is\phi_{\beta}/2} e^{-isk_R x_{\beta}}
      d_{ns}
           +H.c \right]\ \ .  \nonumber
\end{eqnarray}

\subsection{Rashba SO interaction (II)}

Now we second quantize the second term of the Rashba interaction
Eq.(\ref{Rashba2}),
$H_{R2} \equiv -\frac{\alpha(x) \hat{\sigma}_x p_z}{\hbar}$, which can
be accomplished by calculating the matrix elements $\langle
ms'|u(x)^{\dagger} H_{R2} u(x) |ns \rangle  = \langle ms'| H'_{R2}
|ns \rangle $. If $s'=s$, this matrix element is exactly zero. Hence
we only need to calculate the non-diagonal matrix elements, and
they are:
\begin{eqnarray}
 && \hspace{-1.2cm}\langle m\downarrow| H'_{R2} |n \uparrow \rangle
 \nonumber\\
 && = \frac{-\hbar k_R}{m^*}\int d{\bf r} \ e^{-2i k_R x
}\varphi_m({\bf r}) p_z \varphi_n({\bf r})
  \equiv t^{so}_{mn}
\label{hr2a} \\
&&\hspace{-1.2cm}
 \langle n\downarrow| H'_{R2}  |m\uparrow \rangle \nonumber \\
 &&= \frac{-\hbar k_R}{m^*} \int d{\bf r} e^{-2i k_R x
 }\varphi_n({\bf r}) p_z
   \varphi_m({\bf r}) \nonumber \\
 && = \frac{\hbar k_R}{m^*} \int d{\bf r} e^{-2i k_R x
 }\varphi_m({\bf r}) p_z
   \varphi_n({\bf r})
    = -t^{so}_{mn}
\label{hr2b} \\
&&\hspace{-1.2cm}
 \langle n\uparrow| H'_{R2}  |m\downarrow \rangle
 = t^{so*}_{mn}
\label{hr2c} \\
&&\hspace{-1.2cm}
 \langle m\uparrow| H'_{R2} |n\downarrow \rangle
  =-t^{so*}_{mn}
\label{hr2d}
\end{eqnarray}

Here (as well as in below) we have assumed that $k_R(x)$ (or
$\alpha(x)$) to be independent of $x$, but even if $k_R(x)$
depends on $x$, all results are completely the same. With the
above matrix elements (\ref{hr2a}-\ref{hr2d}), the second
quantized form of $H_{R2}$ is:
\begin{eqnarray}
 H_{R2} = \sum\limits_{m,n(m<n)} t^{so}_{mn} \left[
     d^{\dagger}_{m\downarrow} d_{n\uparrow} -
     d^{\dagger}_{n\downarrow} d_{m\uparrow} \right]
   +H.c.
\nonumber
\end{eqnarray}
which can be written in a more compact form:
\begin{eqnarray}
 H_{R2} = \sum\limits_{m,n} t^{so}_{mn}
     d^{\dagger}_{m\downarrow} d_{n\uparrow}
   +H.c.
\label{HR2-result}
\end{eqnarray}
where it is important to realize that $t^{so}_{mn} = -t^{so}_{nm}$.
Eq.(\ref{HR2-result}) is another main result of this paper.

Some general characteristics of Eq.(\ref{HR2-result}) are in
order. (i) The property $t^{so}_{mn} = -t^{so}_{nm}$ for the
matrix elements originate from the time-reversal invariance of the
original real space Rashba Hamiltonian.\cite{addnote} Using this
property, we can exactly prove that the second quantized form of
$H_{R2}$, Eq.(\ref{HR2-result}), indeed satisfies the
time-reversal invariance.\cite{addnote} (ii) If $n=m$, we have
$t^{so}_{nn}=-t^{so}_{nn}$ hence $t^{so}_{nn}$ must vanish. This
means that the Rashba SO interaction can not induce any
intra-level spin flip, {\it i.e.} it can not give rise to a
transition $(n\uparrow)\rightarrow (n\downarrow)$ in which the
level index $n$ is the same. Therefore, the SO interaction
(\ref{HR2-result}) is fundamentally different from that of an
external magnetic field: magnetic field can cause intra-level
spin-flip, provide a Zeeman energy that relieves spin-degeneracy,
and induce spin-polarization in an isolated QD (see subsection
II-E below). (iii) The Rashba SO interaction (\ref{HR2-result})
can cause spin flips between different energy levels. This
inter-level spin-flip coupling is similar to the intersubband
mixing in real space which has been studied in previous
works\cite{ref10}. Despite of the inter-level spin flips, the
system is still at least two-fold degenerate for any eigenstates
because $t^{so}_{mn} = -t^{so}_{nm}$. This guarantees that at
equilibrium an isolated QD has no spin polarization. In the
Appendix, the general properties of Eq.(\ref{HR2-result}) are
further discussed. (iv) In fact, because all spin-orbit couplings
satisfy time-reversal invariance, the above properties and matrix
elements $t_{mn} = -t_{nm}$ must hold true in general. In this
regard, we note that there existed papers where $t_{mn}=t_{nm}$
and intra-level spin-flips are allowed: these effects cannot come
from SO interactions as some times claimed.

Let's estimate the value of $t^{so}_{nm}$. Consider a square QD
with linear size $W$. The eigenstates are $\varphi_n({\bf r})=
\frac{2}{W}\sin\frac{n_x \pi x}{W} \sin \frac{n_z \pi z}{W}$,
hence $t^{so}_{mn}$ can be easily calculated from
Eqs.(\ref{hr2a}). For parameters $W=100nm$, $\alpha=3\times
10^{-11} eVm$ and $m^*=0.036m_e$, the intradot level spacing
$\Delta\epsilon \approx \frac{\hbar^2\pi^2}{2m^* W^2} \approx
1meV$. This is to be compared with a rough estimate of
$|t^{so}_{mn}| \sim \frac{\hbar^2 k_R}{m^* W} = \frac{\alpha}{W}
\sim 0.3meV$.

\subsection{Electron-electron Coulomb interaction}

In order to second quantize the Rashba SO interaction, we have
introduced an unitary transformation defined by Eqs.(\ref{U1}).
Does this transformation affect the familiar second quantized form
of the e-e interaction? Here we show it does not.

Starting from the two-body e-e interaction in real space:
\begin{eqnarray}
 H_I({\bf r}_1,{\bf r}_2,...{\bf r}_N)
  =\sum\limits_{i,j(i\not=j)} \frac{e^2}{2|{\bf r}_i-{\bf r}_j|^2}\ ,
\nonumber
\end{eqnarray}
we apply the unitary transformation and the new Hamiltonian $H'_I$ is:
\begin{eqnarray}
  H'_I & = & \sum\limits_{i,j(i\not=j)} u^{\dagger}(x_i) u^{\dagger}(x_j)
               \frac{e^2}{2|{\bf r}_i-{\bf r}_j|^2}
              u(x_j) u(x_i) \nonumber\\
  & = &
  \sum\limits_{i,j(i\not=j)} \frac{e^2}{2|{\bf r}_i-{\bf r}_j|^2}\ .
\nonumber
\end{eqnarray}
This means $H'_I =H_I$ and the unitary transformation does not
affect the form of the e-e interaction. We therefore can directly
write down the second quantized e-e interaction in its familiar
form,
\begin{eqnarray}
  H_I=\sum\limits_{ns,ms'(ns\not=ms')}U_{ns,ms'}
  d^{\dagger}_{ns} d_{ns}
  d^{\dagger}_{ms'} d_{ms'}
\label{Hee1}
\end{eqnarray}
where the matrix element $U_{ns,ms'}$ is:
\begin{eqnarray}
  U_{ns,ms'} = \langle ns,ms'|\frac{e^2}{2|{\bf r}_1-{\bf r}_2|^2}
          |ms',ns \rangle\ \ .
\nonumber
\end{eqnarray}

\subsection{External magnetic field}

The unitary transformations Eqs.(\ref{U1}) does affect the second
quantized form of the external magnetic fields $\hat{\bf
\sigma}\bullet {\bf B}$. Consider an arbitrary external magnetic
field ${\bf B}=(B_x,B_y,B_z)$ where $B_{x/y/z}$ is projection in
the $x/y/z$ direction.

First, we investigate the z-direction element $B_z$. Under the
unitary transformation Eqs.(\ref{U1}), the term $\hat{\sigma}_z
B_z$ changes to:
\begin{equation}
 u^{\dagger}(x)\hat{\sigma}_z B_z u(x) =
 e^{i\hat{\sigma}_z k_R x} \hat{\sigma}_z B_z e^{-i\hat{\sigma}_z k_R x}
 =\hat{\sigma}_z B_z
\nonumber
\end{equation}
which means $\hat{\sigma}_z B_z$ does not change under the unitary
transformation. Therefore its second quantized form is still
\begin{eqnarray}
  \sum\limits_{ns} sB_z d^{\dagger}_{ns} d_{ns}\ \ .
\label{Bz1}
\end{eqnarray}

Second, we investigate the x-direction element $B_x$, {\it i.e}
the term $\hat{\sigma}_x B_x$ in the Hamiltonian. After the
unitary transformation Eqs.(\ref{U1}), $u^{\dagger}(x)
(\hat{\sigma}_x B_x) u(x) \neq \hat{\sigma}_x B_x$ so that it is
affected by the transformation. The matrix elements $\langle
ms'|u^{\dagger}(x) \hat{\sigma}_x B_x u(x) |ns \rangle $ are found
to be:
\begin{eqnarray}
\langle m\uparrow|u^{\dagger} \hat{\sigma}_x B_x u |n\uparrow \rangle   &=&
\langle m\downarrow|u^{\dagger} \hat{\sigma}_x B_x u |n\downarrow \rangle
= 0  \label{Bx1} \\
 \langle m\downarrow|u^{\dagger} \hat{\sigma}_x B_x u |n\uparrow \rangle
 & =& \int d{\bf r} e^{-i 2 k_R x}\varphi_m({\bf r})
             \varphi_n({\bf r})  B_x  \nonumber \\
 & \equiv & t^{B}_{mn}B_x   \\
\langle n\downarrow|u^{\dagger} \hat{\sigma}_x B_x u |m\uparrow \rangle
&=& t^{B}_{mn} B_x   \label{Bx3}  \\
\langle n\uparrow|u^{\dagger} \hat{\sigma}_x B_x u |m\downarrow \rangle   &=&
\langle m\uparrow|u^{\dagger} \hat{\sigma}_x B_x u |n\downarrow \rangle
\nonumber \\
& = & t^{B*}_{mn}B_x  \label{Bx4}
\end{eqnarray}
Hence, the second quantized form of $\hat{\sigma}_x B_x$ is:
\begin{eqnarray}
&& \sum\limits_{m,n(n<m)} B_x t^B_{mn} \left[
     d^{\dagger}_{m\downarrow} d_{n\uparrow} +
     d^{\dagger}_{n\downarrow} d_{m\uparrow} \right] \nonumber \\
&&  +\sum\limits_{n}  B_x t^B_{nn}
 d^{\dagger}_{n\downarrow} d_{n\uparrow} +H.c. \nonumber
\end{eqnarray}
or it can be written in a more compact form,
\begin{eqnarray}
 \sum\limits_{m,n} B_x t^B_{mn}
     d^{\dagger}_{m\downarrow} d_{n\uparrow}
   +H.c.
\label{Bx2}
\end{eqnarray}
with $t^B_{mn} = t^B_{nm}$. Note that the form of Eq.(\ref{Bx2})
is very similar to the Rashba term $H_{R2}$,
Eq.(\ref{HR2-result}), but there exists an essential difference.
Namely for magnetic field in the x-direction, $t^B_{mn} =
t^B_{nm}$ in Eq.(\ref{Bx2}); while for the Rashba term,
$t^{so}_{mn} = - t^{so}_{nm}$ in Eq.(\ref{HR2-result}). We
emphasize that this is an essential difference because of two
reasons: (i) The magnetic field term destroys time-reversal
invariance, it provides a Zeeman energy that breaks the spin
degeneracy of the energy levels, and it can induce a
spin-polarization in equilibrium. In contrary, the Rashba term
$H_{R2}$ satisfies time-reversal invariance and maintains the two
degeneracies. (ii) When $n=m$, $t^B_{nn}$ can be non-zero so that
intra-level spin flips are possible. Furthermore, the $t^B_{nn}$
term is usually the largest term in the sum of Eq.(\ref{Bx2}),
{\it e.g.} $t^B_{mn} =\delta_{mn}$ at $k_R=0$. But for Rashba
interaction (\ref{HR2-result}), $t^{so}_{nn}$ must vanish as
discussed before so that it cannot cause intra-level spin flip. We
therefore comment that interactions of the following form, which
has been used in some previous literature,
\[
td^{\dagger}_{\downarrow}d_{\uparrow}
 + td^{\dagger}_{\uparrow}d_{\downarrow}
\]
does not represent SO interaction. Rather, it describes a magnetic field
pointing to the x-direction.

In order to estimate the value of $t^B_{nm}$, we consider a
rectangular QD with length $L$ and width $W$. $t^B_{mn}$ can be
obtained as:
\begin{eqnarray}
  t^B_{mn} = 2\delta_{m_z,n_z} \int_0^1 dx e^{-2ik_R L x} \sin m_x \pi x
  \sin n_x \pi x
\nonumber
\end{eqnarray}
Fig.4 plots numerical results for $t^B_{m1}$ versus parameter $k_RL $
obtained this way. As $k_RL$ increases, more $t^B_{m1}$ are in action. If
parameters $\alpha=2*10^{-11} eVm$ and $L =100nm$, we have $k_RL \approx 1$
for $m^* = 0.036 m_e$. For this $k_R L$ value, only $t^B_{11}$ and
$t^B_{21}$ are significant.

Finally, the second quantization of the $B_y$ term is completely the same
as that for the x-direction, hence its second quantized form
is the same as Eq.(\ref{Bx2}).

\subsection{Brief summary}

Collecting all the pieces of second quantization which we have carried out
in this section, for a device in the form of Metal-QD-Metal where there
exists Rashba SO and e-e interactions in the QD, the Metal leads
are magnetic material, and there exists an external magnetic field ${\bf B}$,
Hamiltonian (\ref{Hdevice}) becomes:
\begin{eqnarray}
H  = H_{QD} + \sum\limits_{\beta=L,R}H_{\beta} + H_T
\label{Sum1}
\end{eqnarray}
where
\begin{eqnarray}
H_{QD} & = &
\sum\limits_{n,s}(\epsilon_{n}+sB_z) d^{\dagger}_{ns} d_{ns} \nonumber \\
& + &\sum\limits_{ns,ms'(ns\not=ms')}U_{ns,ms'}
    \hat{n}_{ns} \hat{n}_{ms'} \nonumber \\
& +& \sum\limits_{m,n} \left[ t^{so}_{mn}
d^{\dagger}_{m\downarrow} d_{n\uparrow} +  B_x t^B_{mn}
     d^{\dagger}_{m\downarrow} d_{n\uparrow}
   +H.c. \right]
\label{Sum2} \\
  H_{\beta} & = & \sum\limits_{ k,s} (\epsilon_{k\beta} +sM_{\beta})
     a^{\dagger}_{k\beta s} a_{k\beta s}
\label{Sum3} \\
  H_T & =&  \sum\limits_{k,n,s,\beta}
      \left[t_{k\beta n}
   (\cos \frac{\theta_{\beta}}{2} a^{\dagger}_{k\beta s}
    -s\sin \frac{\theta_{\beta}}{2} a^{\dagger}_{k\beta\bar{s}})
    \right. \nonumber \\
    && \left.\times
    e^{is\phi_{\beta}/2} e^{-isk_R x_{\beta}}
      d_{ns}
           +H.c \right]
\label{Sum4}
\end{eqnarray}
where $t^{so}_{mn} = -t^{so}_{nm}$ and $t^B_{mn} = t^B_{nm}$. This
Hamiltonian is the central result of this paper. The Rashba SO
interaction causes two effects. (i) It gives rise to an extra
phase factor $-sk_R x_{\beta}$ in the hopping matrix element
between the leads and the QD. Note that this phase factor is
dependent on the electronic spin $s$, and it is essential
different from the usual phase factor due to a magnetic flux
which is independent of $s$. (ii) The Rashba SO interaction
causes an inter-level spin-flip term with the strength
$t^{so}_{mn}$ and it cannot cause intra-level spin flips.
Time-reversal invariance is maintained by the SO interaction which
is essential different from the effect of an external magnetic
field.

\section{Example: transport properties of an AB ring with Rashba SO
interaction}
\label{sec4}

As an example of applying the second quantized Hamiltonian
Eqs.(\ref{Sum1}-\ref{Sum4}), we now investigate quantum transport
properties of a modified AB ring shown in Fig.5. A QD sits on one
arm of the ring and Rashba SO interaction exists inside the QD. No
Rashba interaction exists on the other arm of the ring. The ring
is connected to the outside world by two normal metal leads. AB
rings with an embedded QD have been studied by many previous
works\cite{ref29,ref30,ref31,ref32}. Some interesting phenomena
such as Fano resonance\cite{ref30,ref31,ref32} have been
discovered in such a device. The effect of Rashba interaction has
not been studied so far, and we have found that it leads to
interesting transport behavior. In particular, a substantial
spin-polarized current or conductance is induced by a combined
effect of the Rashba SO interaction and a magnetic flux $\phi$
threading through the AB ring. The direction of spin-polarization
and its strength are easily controllable by $\phi$ or by a gate
voltage.

The Hamiltonian of our AB ring (Fig.5) can be written using various pieces
of the general Hamiltonian Eqs.(\ref{Sum1}-\ref{Sum4}):
\begin{eqnarray}
H &=& \sum\limits_{k,s,\beta(\beta=L,R)}
       \epsilon_{\beta k} a^{\dagger}_{\beta k s} a_{\beta k s}
     + \sum\limits_{s} \epsilon_d d^{\dagger}_{s} d_s
      + U d^{\dagger}_{\uparrow}d_{\uparrow}
         d^{\dagger}_{\downarrow}d_{\downarrow} \nonumber\\
  & + & \sum\limits_{k,s} t_{LR} \left[
        a^{\dagger}_{Lks} a_{Rks}    + a^{\dagger}_{Rks} a_{Lks}
        \right] \nonumber \\
  & + & \sum\limits_{k,s} \left[
       t_{Ld} a^{\dagger}_{Lks} d_s +
        t_{Rd} e^{-i s k_R L} e^{i\phi} a^{\dagger}_{Rks} d_{s} \right]
        +H.c.
\label{Hring}
\end{eqnarray}
As discussed previously, the physical meaning of each term is
clear: the first term describes the normal metal leads; the second
term is for the QD which has a single energy level with spin index
$s$; the third term is the intradot e-e Coulomb interaction with a
constant strength $U$; the fourth term is for the arm of the ring
without the QD; and the fifth term is the coupling between the
leads and the QD. Due to the Rashba SO interaction, according to
Eq.(\ref{Sum4}) there is a spin-dependent phase factor $-sk_RL$ in
the hopping matrix element $t_{Rd}$ on the fifth term. Since we
only consider one level in the QD, the inter-level spin-flip term
of Eq.(\ref{Sum2}) does not appear here. This is equivalent to
neglecting the intersubband mixing as in some previous
work\cite{ref6}. We emphasize that both the e-e Coulomb
interaction and Rashba SO interaction are considered, different
from previous studies of the Rashba SO interaction where e-e
interaction was neglected. Indeed, our second quantized
Hamiltonian in the spectral space Eqs.(\ref{Sum1}-\ref{Sum4})
allows us to consider both effects together. Finally, the magnetic
flux $\Phi$ threading the AB ring gives rise to a familiar
spin-independent phase factor $\phi =2\pi \Phi/\Phi_0$ in the
matrix element $t_{Rd}$.

The quantum transport problem described by Hamiltonian
(\ref{Hring}) can be solved by standard many-body techniques. In
the following we calculate charge current using the standard
Keldysh non-equilibrium Green's function method. Following
Ref.\onlinecite{ref18}, the charge current flowing from the left
lead into the AB ring, contributed by spin-up or spin-down
electrons, can be derived as:
\begin{eqnarray}
  I_s = \frac{2e}{\hbar} \int \frac{d\omega}{2\pi} Re
         \left[ t_{Ld} G^<_{dLs}(\omega)
              + t_{LR} G^<_{RLs}(\omega) \right]
\label{curr1}
\end{eqnarray}
where Keldysh Green's function $G^<(\omega)$ is the Fourier
transform of $G^<(t)$ whose definition is:
\begin{eqnarray}
  G^<_{\beta\beta' s}(t) & \equiv &
   i\langle \sum\limits_{k'} a^{\dagger}_{k'\beta' s}(0)
    \sum\limits_{k}  a_{k\beta s}(t)  \rangle  \nonumber \\
  G^<_{\beta d s}(t) & \equiv &
   i\langle  d^{\dagger}_{ s}(0)
     \sum\limits_{k}  a_{k\beta s}(t)  \rangle  \nonumber \\
  G^<_{d d s}(t) & \equiv &
   i\langle  d^{\dagger}_{ s}(0)
      d_{s}(t)  \rangle\ \ .
\label{Gless}
\end{eqnarray}

To solve $G^<$, we first calculate the retarded Green functions ${\bf G}_s^r$
using the Dyson equation:
\begin{equation}
{\bf G}_s^r = {\bf g}_s^r + {\bf g}_s^r {\bf \Sigma}_s^r {\bf G}_s^r
\label{Dyson1}
\end{equation}
and the Green's function ${\bf G}_s^r$ is a $3\times 3$ matrix defined as:
\begin{equation}
 {\bf G}_s^r \equiv \left( \begin{array}{lll}
   G^r_{LLs} & G^r_{LRs} & G^r_{Lds} \\
   G^r_{RLs} & G^r_{RRs} & G^r_{Rds} \\
   G^r_{dLs} & G^r_{dRs} & G^r_{dds}
   \end{array}
  \right)
\label{Gr1}
\end{equation}
In Eq.(\ref{Dyson1}), ${\bf g}_s^r$ is the Green's function of the system
without coupling between the leads and the QD ({\it i.e.} when
$t_{LR}=t_{Ld}=t_{Rd}=0$). It can be obtained exactly as:
\begin{equation}
 {\bf g}_s^r(\omega) \equiv \left( \begin{array}{lll}
   -i\pi \rho & 0 & 0 \\
   0 & -i\pi \rho & 0 \\
   0 & 0 & g^r_{dds}(\omega)
   \end{array}
  \right)
\label{gsr}
\end{equation}
where $g_{dds}^r(\omega)= \frac{\omega-\epsilon_{d}-U+Un_{\bar{s}}
}{(\omega-\epsilon_{d})(\omega-\epsilon_{d}-U) }$ and
$n_{\bar{s}}$ is the intradot electron occupation number at
state $\bar{s}$. $\rho$ in Eq.(\ref{gsr}) is the density of state of the
leads. The self-energy ${\bf \Sigma}^r_s(\omega)$ in
Eq.(\ref{Dyson1}) is:\cite{ref33}
\begin{equation}
 {\bf \Sigma}_s^r(\omega) \equiv \left( \begin{array}{lll}
   0  & t_{LR} & t_{Ld} \\
   t_{LR}^* & 0 &  \tilde{t}_{Rds}\\
   t^*_{Ld} & \tilde{t}^*_{Rds} & 0
   \end{array}
  \right)
\label{Sigma1}
\end{equation}
where $\tilde{t}_{Rds} =t_{Rd}e^{-isk_RL}e^{i\phi}$. Using
Eqs.(\ref{gsr},\ref{Sigma1}), ${\bf G}^r_s$ can easy be obtained by solving
the Dyson's equation (\ref{Dyson1}) as:
${\bf G}^r_s = ({\bf g}_s^{r-1} - {\bf \Sigma}^r_s )^{-1} $.

After solving ${\bf G}^r_s(\omega)$, the Keldysh Green's function
${\bf G}^<_s(\omega)$ can be obtained straightforwardly from the standard
Keldysh equation:
\begin{eqnarray}
{\bf G}^<_s & = & ({\bf 1} +{\bf G}^r_s {\bf \Sigma}^r_s) {\bf g}^<_s
              ({\bf 1} + {\bf \Sigma}^a_s {\bf G}^a_s)
            +{\bf G}^r_s {\bf \Sigma}^<_s {\bf G}^a_s \nonumber \\
 & = & {\bf G}^r_s {\bf g}^{r-1}_s {\bf g}^<_s
               {\bf g}^{a-1}_s {\bf G}^a_s
            +{\bf G}^r_s {\bf \Sigma}^<_s {\bf G}^a_s
\label{Keldysh1}
\end{eqnarray}
For our present case, ${\bf \Sigma}^<_s =0$ and ${\bf g}^{r-1}_s
{\bf g}^<_s {\bf g}^{a-1}_s $ is diagonal with ${\bf
g}^{r-1}_{\beta\beta s} {\bf g}^<_{\beta\beta s} {\bf
g}^{a-1}_{\beta\beta s} =2i f_{\beta}(\omega)/\pi\rho$
($\beta=L,R$) and ${\bf g}^{r-1}_{dd s} {\bf g}^<_{dd s} {\bf
g}^{a-1}_{dd s} = 0$, where $f_{\beta}(\omega)
=1/(e^{(\omega-\mu_{\beta})/k_B{\cal T} }+1)$ is the Fermi
distribution function in lead $\beta$. As the last step, the
intradot electron occupation number $n_{s}$ needs to be solved
self-consistently with the self-consistent equation $n_{s} =-i\int
\frac{d\omega}{2\pi} G^<_{dds}(\omega)$.

In the following we present numerical results. Fig.6 shows the total
linear conductance $G = \sum_s \frac{d I_s}{dV}$ versus the
intradot level position $\epsilon_d$ at zero magnetic flux ($\phi=0$)
but with different Rashba interaction strength $k_R L$:
$k_R L = 0$ (solid), $\pi/4$ (dashed), $\pi/2$ (dotted), $3\pi/4$
(dash-dotted), and $\pi$ (dash-dot-dotted). The curves are dominated by
two Coulomb peaks at $\epsilon_d =0$ and $\epsilon_d =-U$. When there is no
Rashba SO interaction ({\it i.e.} $k_R L =0$, solid curve), these two peaks
show a typical Fano resonance shape due to the interference of electrons
passing the two arms of the AB ring, in agreement with previous theoretical
and experimental studies\cite{ref29,ref30}. It is interesting to discover
that this Fano resonance can be strongly affected by the Rashba SO
interaction.
Increasing the Rashba parameter $k_R L$ from $0$, the Fano resonance is
decreased and it can completely disappear at $k_R L = \pi/2$ (dotted curve).
Further increasing $k_R L$, the Fano resonance rises up again but with an
oppositive Fano factor, for example at $k_R L = \pi$ (dash-dot-dotted curve).

In order to understand these results, we investigate the interference term
of total transmission probability, which is approximatively
proportional to $\sim \sum_s \cos(\Delta\theta +sk_RL) $. Here
$\Delta \theta$ is the phase difference of the transmission amplitude through
the two arms, and it varies from $0$ to $\pi/2$ and finally to $\pi$ as
$\epsilon_d$ is moved from $-\infty$ to $0$ and finally to $\infty$.
This clearly shows that the total transmission probability is indeed having
a Fano asymmetric resonance shape when $k_R L = 0$ or $\pi$. On the other
hand, it is symmetric at $k_R L =\pi/2$ or $3\pi/2$.  Hence the Rashba SO
interaction can alter the Fano resonance shape in substantial ways.

Next, in the three panels of Fig.7 and Fig.8, we plot conductance
$G_s$ and spin polarization $\eta \equiv (G_{\uparrow}
-G_{\downarrow})/(G_{\uparrow}+G_{\downarrow})$ versus magnetic
flux $\phi$ for three values of $\epsilon_d=1,0,-1$, respectively.
These values of $\epsilon_d$ are near the right Coulomb peak of
Fig.6. In Fig.7, the thick curves are for $G_{\uparrow}$ and thin
curves for $G_{\downarrow}$, the solid, dashed, and dotted curves
correspond different values of the Rashba parameter $k_R L =0$,
$\pi/4$, and $\pi/2$. Figs.7,8 clearly show that if either one of
the two parameters ($\phi$ and $k_R L$) vanishes, the transport
current has no spin polarization so that $\eta=0$ and
$G_{\uparrow} = G_{\downarrow}$. However, when both parameters are
nonzero, a substantial spin-polarized conductance is found and
$\eta$ can be as large as $90\%$ for the given set of system
parameters (Figs.8a,c).

Importantly, in the present device the direction of
spin-polarization and its strength ($\eta$) are easily
controllable by varying system parameters which are experimentally
accessible. (i) By varying the magnetic flux $\phi$: when $\phi$
is tuned from $-\pi/2$ to $\pi/2$ (or from $\pi/2$ to $3\pi/2$),
the polarization $\eta$ strongly varies from a large positive
value to a negative value or vice versa. (ii) By varying the
intradot level $\epsilon_d$ using a gate voltage: when
$\epsilon_d$ is moved from one side to another side of a Coulomb
peak, the polarization $\eta$ can be tuned from its largest
positive value to its largest negative value or vice versa.
Numerically we found that one only needs to change $\epsilon_d$ by
a small amount to see the polarization change, namely a few
half-widths $\Gamma$ of the Coulomb peak (parameters used in
Figs.7,8 correspond to $\Gamma \equiv 2\pi \rho |t_{\beta d}|^2
\approx 1$). This means that in an experiment one only needs to
slightly vary the gate voltage to change $\eta$ from $1$ to $-1$
or vice versa. Furthermore, we note that when polarization $\eta$
reaches its largest value, the conductance itself is still large,
{\it e.g.} $G_{\uparrow}$ or $G_{\downarrow}$ can exceed over
$0.8e^2/h$ (see Fig.7a).

Finally, we estimate if the parameter $k_R L$ can reach a value of $\pi/2$
in the present experimental technology so that the above theoretical
predictions can be observed experimentally. Assuming the Rashba SO interaction
strength $\alpha\sim 3\times 10^{-11}eVm $, which is
the reported value for some semiconductors\cite{ref10,ref23,ref24},
$k_R = m^* \alpha /\hbar^2 \approx 0.015/nm$ for $m^* = 0.036m_e$.
Then, if the length of the QD is the typical value $100nm$, $k_R L \approx
1.5$. Therefore we conclude that $k_R L$ can reach a value $\sim \pi/2$ or
larger experimentally.

\section{Conclusion}

In this paper we have derived a second quantized Hamiltonian in
{\it spectral space} for a general device structure of
Metal-QD-Metal configuration, including the spin-orbital and e-e
interactions. In other words, we extended the standard Anderson
Hamiltonian to the case where the central device region has a
(Rashba) spin-orbital interaction. We discovered that the Rashba
SO interaction causes two changes: (i) It gives rise to an extra
spin dependent phase factor $-sk_R x_{\beta}$ in the coupling
matrix elements between the leads and the quantum dot. (ii) The
Rashba SO interaction causes an inter-level spin-flip term with
strength $t^{so}_{mn}= -t^{so}_{nm}$, and it cannot cause any
intra-level spin flips.

The spectral form of the Hamiltonian is very important as it
allows the analysis of many complicated quantum transport problems
involving SO and e-e interactions, by using the well established
many-body Green's function theoretical techniques such as the
Keldysh nonequilibrium Green's function formalism. On the other
hand, it would be much more difficult to carry out similar
investigations using a real space Hamiltonian, especially if e-e
interactions are present.

As an example, we investigated quantum transport properties of a
AB ring in which a QD having Rashba SO and e-e interactions is
embedded in one arm of the ring. A substantial spin-polarized
current or conductance emerges in this device due to a combined
effect of a magnetic flux and the Rashba SO interaction. In
particular, the direction of the spin polarization and the
strength $\eta$ can be easily controlled by a number of
experimentally accessible parameters.

\section*{Acknowledgments} We gratefully acknowledge financial support
from the Chinese Academy of Sciences and NSF-China under Grant No.
90303016 and 10474125 (Q.F.S.); RGC grant from the SAR Government
of Hong Kong under grant number HKU 7044/04P (J.W.); NSERC of
Canada, FQRNT of Qu\'{e}bec and Canadian Institute of Advanced
Research (H.G). Q.F.S. gratefully acknowledge Professor X.C. Xie
for helpful discussions on the general physics of SO interaction.

\section*{Appendix}

In this appendix, we collect some general properties of the spin-orbit
coupling. Although these properties should be well-known before\cite{ref10},
we believe it is useful to put them in a form that is easily accessible.
In addition, these properties hold for all spin-orbit interactions
including the Rashba SO interaction.

(1). The SO interaction Hamiltonian $H_{so}$, Eq.(\ref{Hso1}),
satisfies time-reversal invariance. In other words, $H_{so}$
commutes with the time-reversal operator $T=-i\hat{\sigma}_y K$
(where $K$ is the complex-conjugation operator). Indeed, using the
Hamiltonian $H_{so}$ of Eq.(\ref{Hso1}), it is easy to prove
$[T,H_{so}] =0$.

(2). When a system has spin-orbit coupling, each eigenenergy level
is still at least two-fold degenerate, {\it i.e.} the so-called
Kramers degeneracy exists. Briefly this can be proved as follows.
We start from the Hamiltonian $H$ of Eqs.(\ref{Hs1}) but setting
${\bf B}={\bf M}({\bf r})=0$ in (\ref{Hs1}), assume
$\varphi_n({\bf r}_1,{\bf r}_2,...{\bf r}_N)$ is an eigen-state of
$H$ so that $H|\varphi_n \rangle =E_n |\varphi_n \rangle $. Since
$H$ is time-reversal invariant ($TH = HT$), we have $HT|\varphi_n
\rangle =TH|\varphi_n \rangle  =E_n T|\varphi_n \rangle $. Hence
state $T|\varphi_n \rangle $ is also an eigen-state with the same
eigen energy $E_n$ as that of the state $|\varphi_n \rangle $.
Furthermore, one has $\langle \varphi_n|T|\varphi_n \rangle  =
(\varphi_n, T\varphi_n) =(T^2\varphi_n, T\varphi_n) =-\langle
\varphi_n|T|\varphi_n \rangle $, hence $\langle
\varphi_n|T|\varphi_n \rangle =0$. This means state $T|\varphi_n
\rangle $ is orthogonal to $|\varphi_n \rangle $. Therefore,
although spin is no longer a good quantum number when SO
interaction exists, the system is still at least two-fold
degenerate for any of its eigen state.

(3). At equilibrium, any spin-orbit coupling cannot induce a spontaneous spin
polarization. We prove this as follows. Since the system is in equilibrium,
the two-fold degenerate eigen-states $|\varphi_n \rangle $ and
$T|\varphi_n \rangle $ have the same occupation probability $p(E_n)$. Then,
the average of spin polarization at an arbitrary direction $\hat{n}$
can be calculated as:
\begin{eqnarray}
 \langle \hat{\sigma}_{\hat{n}} \rangle  & =&
  \sum\limits_n p(E_n) [ \langle \varphi_n|\hat{\sigma}_{\hat{n}}|\varphi_n
\rangle
  + \langle T\varphi_n|\hat{\sigma}_{\hat{n}}|T\varphi_n \rangle   ]
        \nonumber \\
  & =&
 \sum\limits_n p(E_n) [ \langle \varphi_n|\hat{\sigma}_{\hat{n}}|\varphi_n
\rangle
  + \langle T\varphi_n|-T\hat{\sigma}_{\hat{n}}|\varphi_n \rangle   ]
        \nonumber \\
  & =&
 \sum\limits_n p(E_n) [ \langle \varphi_n|\hat{\sigma}_{\hat{n}}|\varphi_n
\rangle
  - \langle \varphi_n|\hat{\sigma}_{\hat{n}}|\varphi_n \rangle   ]
        \nonumber \\
 & =& 0\ \ .
\nonumber
\end{eqnarray}
Therefore, at equilibrium no spin-orbit coupling can induce a spontaneous
spin polarization in any direction.

\newpage

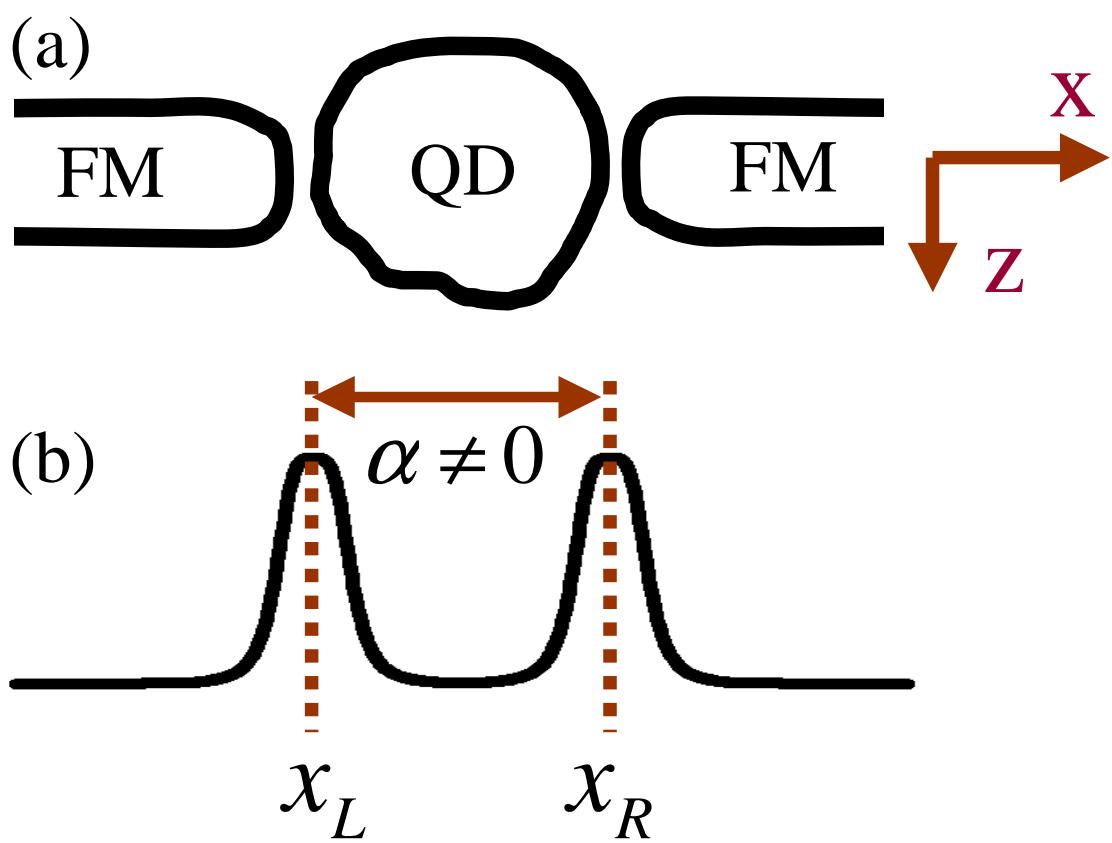
\begin{figure}
\caption{ (a) Schematic diagram for a Metal-QD-Metal device
configuration where the QD is weakly coupled to two ferromagnetic
leads. (b) Schematic diagram for the scattering potential along
the x-direction. The Rashba SO interaction is assumed to only
exist in the central QD region, {\it i.e.} $\alpha =0$ in for
regions with $x<x_L$ and $x> x_R$. } \label{fig:1}
\end{figure}

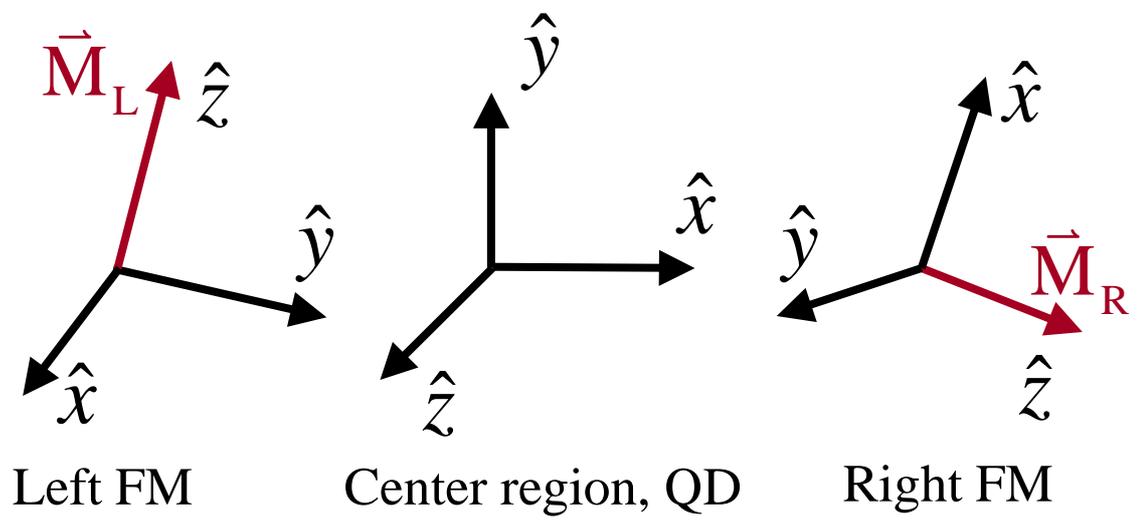
\begin{figure}
\caption{ Schematic diagram for the spin coordinates, {\it i.e.}
the spin-up direction in the left lead, the center region, and the
right lead, respectively. } \label{fig:2}
\end{figure}

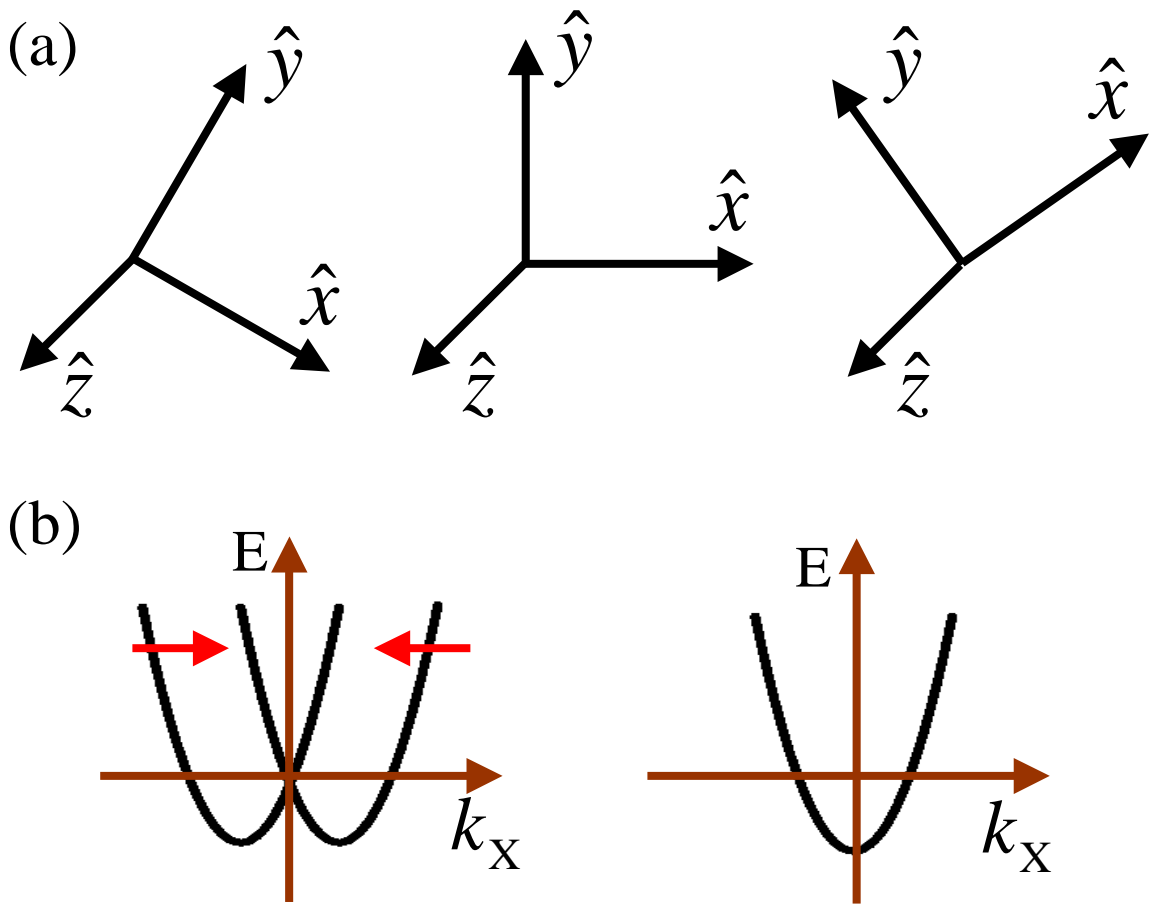
\begin{figure}
\caption{ (a). Schematic diagram for the spin coordinate axis in
different position. Here the $x$ and $y$ spin directions are
rotated along the $x$-axis in space. (b). Schematic diagram for
the dispersion relation before and after the unitary
transformation. } \label{fig:3}
\end{figure}

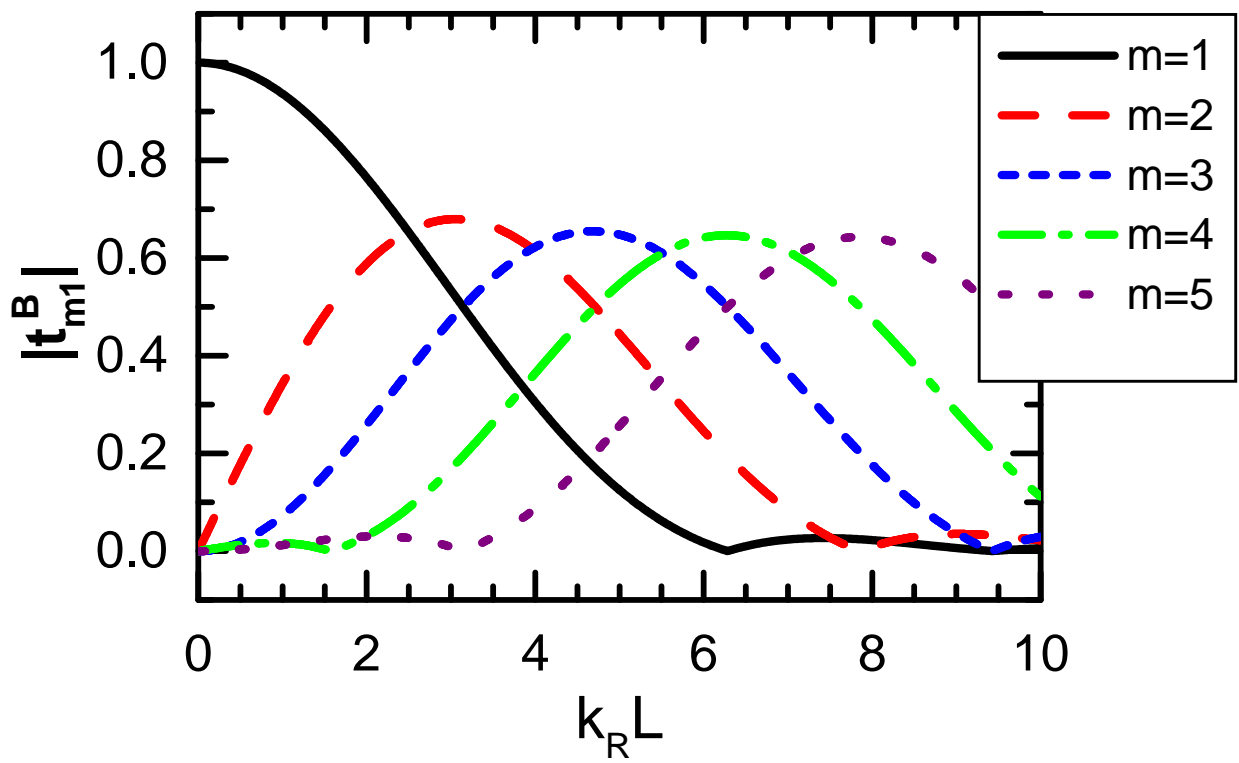
\begin{figure}
\caption{
The spin-flip coupling strength $t^B_{m1}$ versus the Rashba SO interaction
$k_R L$. As $k_RL$ increases, more modes ($m$) are playing a role.
}
\label{fig:4}
\end{figure}

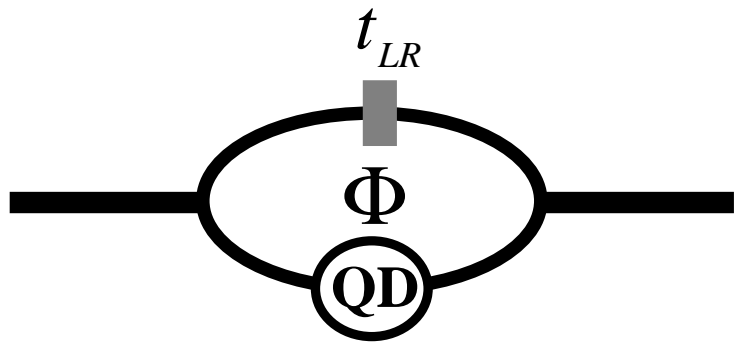
\begin{figure}
\caption{ Schematic diagram for the modified AB ring device: two
normal leads are couplied to the center ring, a magnetic flux
threads the the ring, and a QD is embedded to one arm of the ring.
} \label{fig:5}
\end{figure}

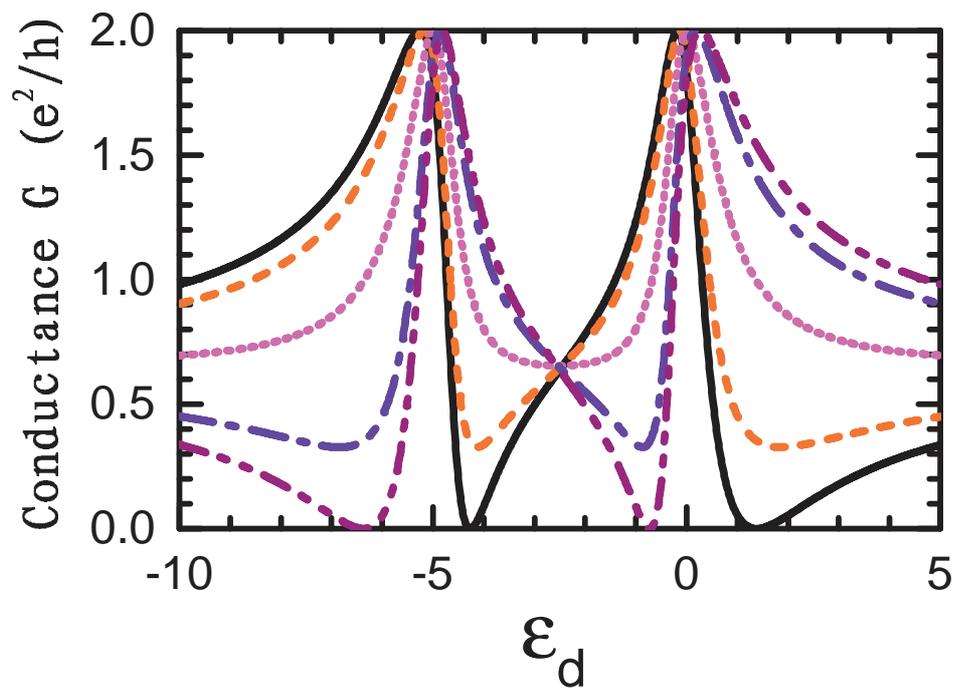
\begin{figure}
\caption{
Conductance $G$ versus the intra-dot level $\epsilon_d$
for $k_R L = 0$ (the black solid curve), $\pi/4$ (the red dashed curve),
$\pi/2$ (the magenta dotted curve), $3\pi/4$ (the blue dash-dotted curve),
and $\pi$ (the purple dash-dot-dotted curve), respectively.
Other parameters are: $t_{Rd}=t_{Ld}=0.4$, $t_{LR} =0.1$,
$\rho_L=\rho_R =1$, $k_B {\cal T}=0.0001$, $U=5$, and $\phi=0$.
}
\label{fig:6}
\end{figure}

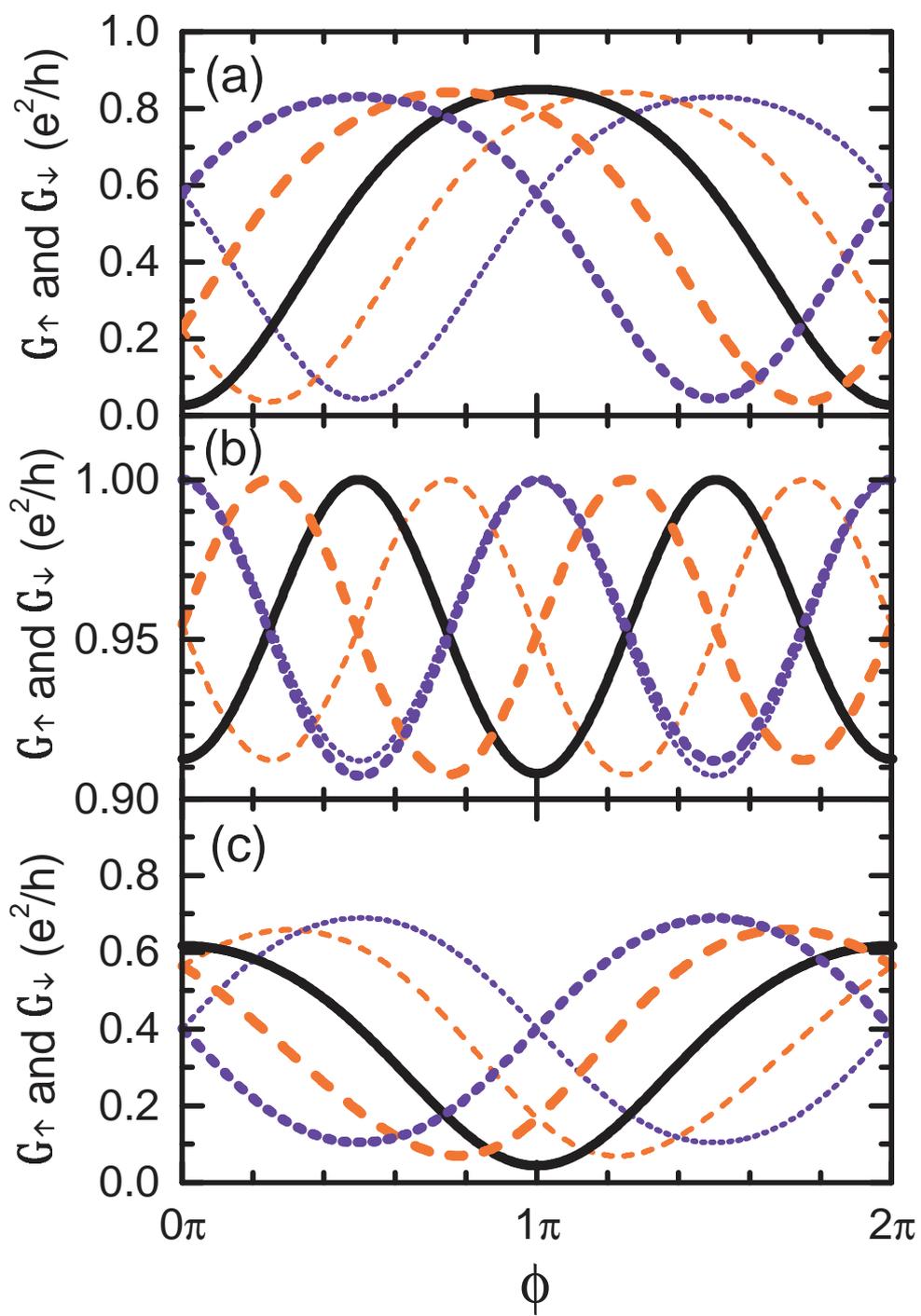
\begin{figure}
\caption{
Conductance $G_{\uparrow}$ and $G_{\downarrow}$ versus
magnetic flux $\phi$ for several intradot levels: (a) $\epsilon_d=1$;
(b) $\epsilon_d=0$; (c) $\epsilon_d=-1$. The solid, dashed, and dotted
curves correspond to $k_R L =0$, $\pi/4$, and $\pi/2$, respectively.
The thick curves are $G_{\uparrow}$ and thin curves are $G_{\downarrow}$.
Other parameters are the same as those of Fig.6.
}
\label{fig:7}
\end{figure}

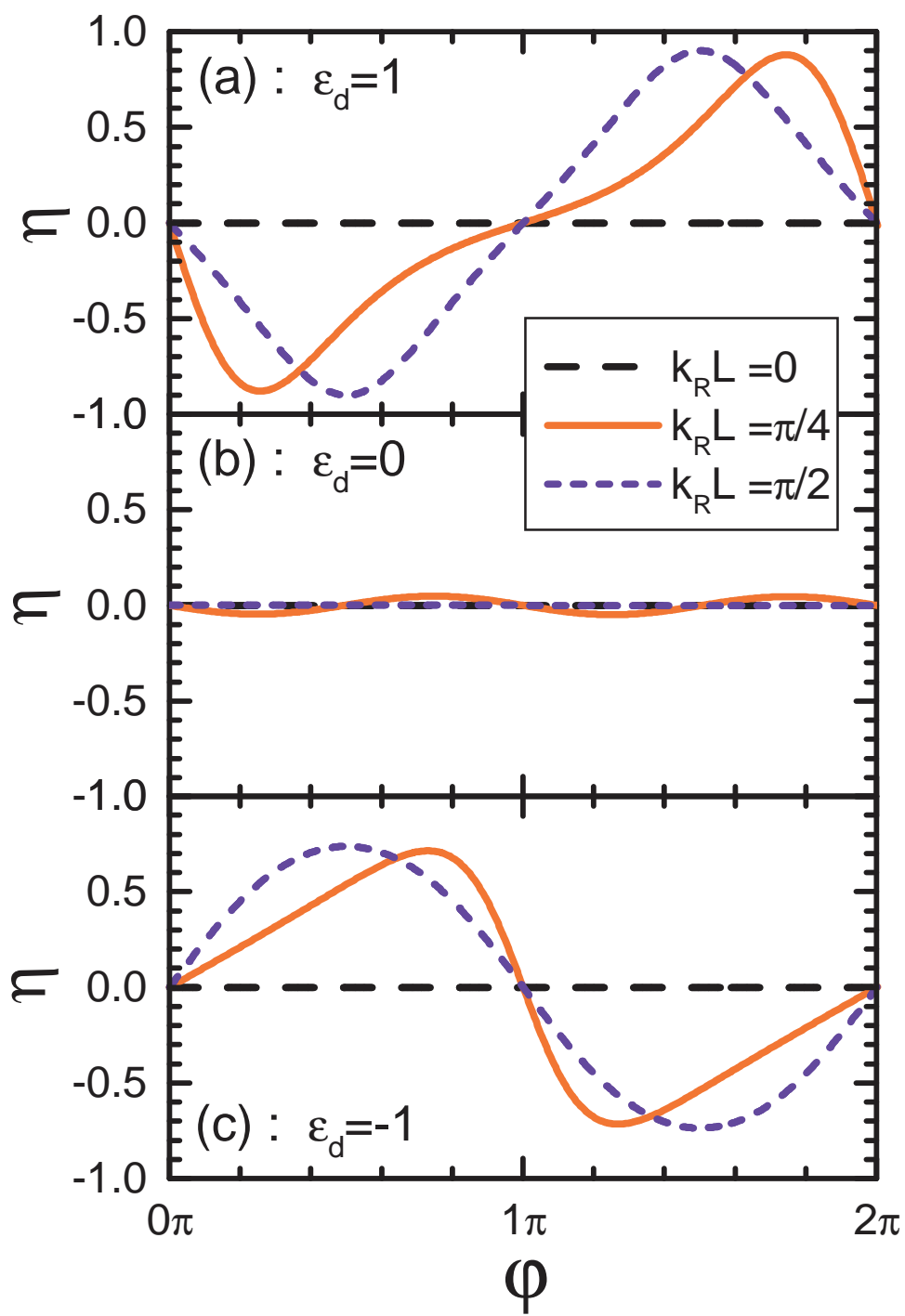
\begin{figure}
\caption{ Spin polarization $\eta$ versus magnetic flux $\phi$ for
$k_R L =0$ (dashed curve), $\pi/4$ (solid curves), and $\pi/2$
(dotted curve), respectively. Other parameters are the same as
those of Fig.6. } \label{fig:8}
\end{figure}

\end{document}